\documentclass[runningheads]{llncs}

\usepackage{booktabs}
\usepackage{colortbl}
\usepackage[T1]{fontenc}
\usepackage{graphics}
\usepackage{xcolor}
\usepackage{xspace}
\usepackage{xurl}

\newcommand{\etal}{et~al.\xspace}
\newcommand{\ie}{i.e.\xspace}
\newcommand{\eg}{e.g.\xspace}
\newcommand{\queen}{{\sc QuEEn}\xspace}
\newcommand{\fakesection}[1]{\vspace{1mm} \noindent {\bf #1}}

\newcommand{\queenexplanation}{%
    {\sc \underline{Qu}}antum {\sc \underline{E}}volved {\sc \underline{E}}{\sc \underline{n}}sembles\xspace
}

\newcommand{\rqone}{%
    What is the difference in fitness between heterogeneous and homogeneous ensembles?\xspace
}

\newcommand{\rqtwo}{%
    How does noise impact the fitness difference between heterogeneous and homogeneous ensembles?\xspace
}

\begin{document}

\title{Improving the Reliability of Quantum Circuits by Evolving Heterogeneous Ensembles}
\titlerunning{Improving Reliability by Evolving Heterogeneous Ensembles}

\author{Owain Parry \and John Clark \and Phil McMinn}

\institute{University of Sheffield \\\email{\{o.b.parry,john.clark,p.mcminn\}@sheffield.ac.uk}}

\maketitle

\begin{abstract}

Quantum computers can perform certain operations exponentially faster than 
classical computers, but designing quantum circuits is challenging.
To that end, researchers used evolutionary algorithms to produce probabilistic 
quantum circuits that give the correct output more often than not for any input.
They can be executed multiple times, with the outputs combined using a classical 
method (such as voting) to produce the final output, effectively creating a 
homogeneous ensemble of circuits (\ie, all identical).
Inspired by $n$-version programming and ensemble learning, we developed a tool 
that uses an evolutionary algorithm to generate heterogeneous ensembles of 
circuits (\ie, all different), named \queen.
We used it to evolve ensembles to solve the Iris classification problem.
When using ideal simulation, we found the performance of heterogeneous ensembles 
to be greater than that of homogeneous ensembles to a statistically significant 
degree.
When using noisy simulation, we still observed a statistically significant 
improvement in the majority of cases.
Our results indicate that evolving heterogeneous ensembles is an effective
strategy for improving the reliability of quantum circuits.
This is particularly relevant in the current NISQ era of quantum computing where 
computers do not yet have good tolerance to quantum noise.


\end{abstract}

\section{Introduction}
\label{sec:introduction}

Quantum computers can efficiently solve certain problems that would be 
intractable for classical computers \cite{Massey2004,Massey2005}.
However, designing quantum circuits is difficult 
\cite{Barnes2019,Massey2004,Zhu2022}, so researchers investigated the 
effectiveness of evolutionary algorithms \cite{Bartz2014} to generate them 
\cite{Barnes2019,Massey2004,Massey2005,Zhu2022}.
To read the output after executing a circuit, the qubits needs to be 
measured, resulting in the collapse of the quantum state into a classical state.
Generally, this is a non-deterministic process.
Massey \etal \cite{Massey2004} evolved a {\it deterministic circuit} for 
adding integers that always gives the correct output for any input.
However, they were unable to evolve deterministic circuits for more challenging 
problems, reasoning that evolutionary algorithms are better suited to finding 
good approximate solutions.
Instead, they evolved {\it probabilistic circuits}, that give the correct output 
more often than an incorrect output for any input.
They may be executed repeatedly, and the outputs combined in some classical way 
to produce the final output (\eg, a vote), effectively equivalent to a 
{\it homogeneous ensemble} of circuits (\ie, all the same).

In software engineering, $n$-version programming is a method used to improve 
software safety \cite{Khoury2012}.
It involves independently producing multiple functionally equivalent programs
from a single specification and executing them in parallel for improved
fault-tolerance through a voting mechanism.
In machine learning, ensemble learning is a state-of-the art solution for 
classification and regression problems \cite{Sagi2018}.
It improves upon the performance of a single predictive model by training a
{\it heterogeneous ensemble} of models (\ie, all different) and combining their 
predictions.
Despite the success of combining the outputs of diverse solutions in software 
engineering and machine learning, there has been no attempt to evolve 
heterogeneous ensembles of quantum circuits to our knowledge.

The Noisy Intermediate-Scale Quantum (NISQ) era refers to the current state of 
quantum computing where computers lack dependable tolerance to quantum noise 
\cite{Chen2023}.
A common manifestation is depolarizing error, where the state of a qubit 
undergoes random perturbations, making circuits less likely to produce the 
correct output \cite{Muqeet2024b,Muqeet2024a}.
As in $n$-version programming, heterogeneous ensembles may provide greater 
reliability in the presence of noise by way of redundancy.
Because the circuits are all different, the impact of noise on each circuit may 
also be different, and so collectively they may be more likely to produce the 
correct final output through a voting mechanism than a homogeneous ensemble.

We developed a tool that uses an evolutionary algorithm to generate 
heterogeneous ensembles of probabilistic quantum circuits, named \queen 
(\queenexplanation).
We used it to evolve ensembles of $3$, $5$, and $7$ circuits to tackle the Iris
classification problem \cite{Chen2005}.
In each case, we found the performance of heterogeneous ensembles to be greater 
than that of homogeneous ensembles of the same size to a statistically 
significant degree when using ideal simulation.
When using noisy simulation, we still observed a statistically significant 
improvement for ensembles of size $5$ and $7$ in the vast majority cases.
These results indicate that our approach produces more reliable ensembles, in 
the sense that they are more likely to produce the correct output even with 
noise.

In summary, the main contributions of this paper are:

\noindent {\bf 1. QuEEn.} We developed a new tool for evolving heterogeneous 
ensembles of quantum circuits (see Section \ref{sec:queen}).

\noindent{\bf 2. Novel Evaluation.} Ours is the first study to compare the 
performance of heterogeneous ensembles of circuits to homogeneous ensembles (see 
Section \ref{sec:evaluation}).

\noindent{\bf 3. Findings and Implications.} Our results indicate that evolving 
heterogeneous ensembles is an effective strategy, potentially paving the way for 
more advanced techniques in quantum circuit evolution (see Section 
\ref{sec:results}).

\section{\bf QuEEn}
\label{sec:queen}

Our tool \queen uses an evolutionary algorithm \cite{Bartz2014} to evolve 
a population of equally-sized heterogeneous ensembles of probabilistic quantum 
circuits that implement a solution to a given problem.
A circuit operates on a quantum register of qubits, following which, one or more 
are measured to populate a classical register of bits that encode the circuit's 
output value.
The tool represents a circuit as a list of quantum gates and their parameters.
It supports two gate types: {\it U gates}, a generic single-qubit rotation gate 
with three angles, and {\it CX gates}, a controlled-not gate.
These two gates form a universal gate set sufficient to express any quantum 
circuit \cite{Barenco1995}.
The tool considers an ensemble to be a set of circuits whose collective output 
value is the most frequent among its members.
In the event of a tie, the ensemble would select uniformly at random among the 
tied circuit output values.
As arguments, \queen takes the number of qubits in the quantum register, a file 
to write the population to, and a file to read test cases from.
A test case represents a specific instance of the problem.
It consists of an initialization circuit to set the quantum register to some 
input value, along with the expected output value to compare against the output 
value of an ensemble.

The tool uses the {\tt qiskit} package \cite{Qiskit2024} to simulate circuit
execution and qubit measurement.
By default, \queen uses ideal simulation that does not consider noise.
Optionally, \queen takes the name of a {\it fake backend}.
These are provided by {\tt qiskit} and approximately reproduce the noise 
measured in specific quantum computers for simulations \cite{Muqeet2024a}.
The state of the classical register after measurement is non-deterministic so 
\queen performs 1,000 repeats of the circuit simulation.
This results in a discrete probability distribution over the possible output 
values.
The tool computes the output distribution for an ensemble based on the 
output distributions of its circuits and the voting scheme described earlier in 
this section.
The {\it fitness} of an ensemble, which measures how well it addresses the 
problem, is its mean probability of producing the expected output value over the 
set of test cases (higher is better).
While more complex fitness functions exist~\cite{Stepney2008}, we selected this
because it has an intuitive interpretation.
Our fitness function does not penalize ensembles based on the number of gates
in its circuits but \queen does enforce a fixed upper bound.
In this initial study, \queen does not explicitly promote circuit diversity 
within ensembles.
However, we plan to implement and evaluate the impact of this as part of future 
work.

\section{Evaluation}
\label{sec:evaluation}

In this section, we describe our methodology for answering our research 
questions regarding evolved ensembles of probabilistic quantum circuits:

\noindent {\bf RQ1.} \rqone

\noindent {\bf RQ2.} \rqtwo

\fakesection{Problem and Test Cases.} 
We used the Iris classification problem \cite{Chen2005} in our evaluation 
because it is non-trivial, yet relatively straight-forward to address and has 
seen extensive use as a benchmark for statistical classification techniques.
It consists of determining the species of an Iris flower from three 
possibilities based on the length and width of its sepal and petal (four 
features).
The corresponding dataset contains 150 examples, 50 for each of the three 
classes, which we converted into test cases for \queen.
For each test case, we generated the appropriate initialization circuit based on
an angle encoding scheme \cite{Weigold2021} that requires one qubit per feature.
This requires a quantum register of four qubits because there are four features.
The expected output for each test case encodes the correct class in two 
classical bits which are to be measured from the first two qubits.
Because there are only three possible classes, one of the four possible output
values represents an ``invalid'' class.
We randomly split the 150 test cases into 100 for evolving ensembles (the 
{\it evolution tests}) and 50 for evaluating the fitness of each ensemble in the 
final population (the {\it evaluation tests}).
This is to avoid any bias in our results caused by overfitting.

\fakesection{Answering our Research Questions.} 
To answer {\bf RQ1}, we invoked \queen four times to evolve ensembles of size 1, 
3, 5, and 7 based on the evolution tests.
Where the ensemble size was 1, we are evolving individual circuits.
In each case, we set the population size to 600 ensembles and the number of 
generations to 5,000.
We selected these values to explore the search space as much as possible within
the constraints of the computational resources available to us.
Once the evolution was complete, we used \queen to evaluate the fitness of each 
ensemble in each of the four final populations based on the evaluation tests.
We did not specify any fake backend during evolution or evaluation, meaning 
\queen always used ideal simulation.
For the final population of individual circuits (ensemble size of 1), we 
additionally used \queen to evaluate the fitness of each circuit as a 
homogeneous ensemble of size 3, 5, and 7.
This represents executing the same circuit several times and applying the voting
mechanism described in Section~\ref{sec:queen}.
For each of the three final populations of heterogeneous ensembles 
(ensemble size of 3, 5, and 7), we compared the fitnesses to those of the 
corresponding homogeneous ensembles using a two-tailed Mann-Whitney $U$ 
hypothesis test.
In each case, the null hypothesis is that the distribution underlying the 
fitnesses of the heterogeneous ensembles is same as that of the homogeneous
ensembles.
In other words, there is no general difference in fitness (mean probability of 
producing the expected output value) between combining the outputs of several 
different circuits and combining the outputs from repeating the same circuit 
several times.
To answer {\bf RQ2}, we repeated the methodology for answering {\bf RQ1} ten 
times, aside from evolving the ensembles again, specifying a different fake 
backend each time.
In each case, this means \queen used noisy simulation when evaluating ensemble 
fitnesses based on the evaluation tests.
We selected the ten fake backends at random from those provided by {\tt qiskit} 
\cite{Qiskit2024}.

\fakesection{Threats to Validity.} 
Evolutionary algorithms are highly randomized and non-deterministic.
Therefore, our results may be unreproducible.
Ideally, we would have mitigated this risk by performing many reruns of the 
ensemble evolution.
However, given the computational complexity of quantum simulation, this was not
possible within the constraints of the resources available to us.
Instead, we mitigated this risk by performing hypothesis testing between 
populations of ensembles, rather than comparing the best individual ensembles 
directly.
All types of hypothesis tests make assumptions about the distributions 
underlying the data.
If these do not hold, the results are likely to be invalid.
To mitigate this risk, we selected the Mann-Whitney $U$ test because it makes
relatively few assumptions compared to other tests, such as Student's $t$-test.

\section{Results}
\label{sec:results}

\begin{table}[t]
    \centering
    \caption{
        \label{tab:table}
        For three values of ensemble size ($n$): the median fitness (Med Fit) of 
        the heterogeneous (Het) and homogeneous (Hom) ensembles, and the p-value 
        ($p$) and rank-biserial correlation effect size ($r$) from the 
        two-tailed Mann-Whitney $U$ test.
        The first row shows the results where \queen used ideal simulation when 
        evaluating ensemble fitnesses.
        The subsequent rows shows the results where \queen used noisy 
        simulation.
    }
    \resizebox{\textwidth}{!}{%
        \begin{tabular}{lrrrrrrrrrrrr}
            \toprule
            & \multicolumn{4}{c}{\boldmath $n=3$} & \multicolumn{4}{c}{\boldmath $n=5$} & \multicolumn{4}{c}{\boldmath $n=7$} \\
            \cmidrule(r){2-5}
            \cmidrule(lr){6-9}
            \cmidrule(l){10-13}
            & \multicolumn{2}{c}{\bf Med Fit} & & & \multicolumn{2}{c}{\bf Med Fit} & & & \multicolumn{2}{c}{\bf Med Fit} \\
            \cmidrule{2-3}
            \cmidrule{6-7}
            \cmidrule{10-11}
            \multicolumn{1}{c}{\bf Backend} & \multicolumn{1}{c}{\bf Het} & \multicolumn{1}{c}{\bf Hom} & \multicolumn{1}{c}{\boldmath $p$} & \multicolumn{1}{c}{\boldmath $r$} & \multicolumn{1}{c}{\bf Het} & \multicolumn{1}{c}{\bf Hom} & \multicolumn{1}{c}{\boldmath $p$} & \multicolumn{1}{c}{\boldmath $r$} & \multicolumn{1}{c}{\bf Het} & \multicolumn{1}{c}{\bf Hom} & \multicolumn{1}{c}{\boldmath $p$} & \multicolumn{1}{c}{\boldmath $r$} \\
            \midrule
            {\it ideal} & 0.782 & 0.757 & <0.001 & 0.932 & 0.920 & 0.775 & <0.001 & 0.997 & 0.920 & 0.783 & <0.001 & 0.987 \\
\rowcolor{gray!20}
boeblingen & 0.705 & 0.724 & <0.001 & -0.895 & 0.864 & 0.758 & <0.001 & 0.990 & 0.841 & 0.773 & <0.001 & 0.957 \\
cairo & 0.754 & 0.746 & <0.001 & 0.803 & 0.895 & 0.770 & <0.001 & 0.997 & 0.891 & 0.780 & <0.001 & 0.981 \\
\rowcolor{gray!20}
casablanca & 0.727 & 0.734 & <0.001 & -0.784 & 0.872 & 0.764 & <0.001 & 0.990 & 0.859 & 0.777 & <0.001 & 0.964 \\
essex & 0.687 & 0.711 & <0.001 & -0.899 & 0.812 & 0.751 & <0.001 & 0.966 & 0.796 & 0.769 & <0.001 & 0.758 \\
\rowcolor{gray!20}
guadalupe & 0.736 & 0.742 & <0.001 & -0.734 & 0.880 & 0.768 & <0.001 & 0.990 & 0.867 & 0.779 & <0.001 & 0.969 \\
kyoto & 0.250 & 0.251 & <0.001 & -0.212 & 0.251 & 0.251 & 0.003 & -0.099 & 0.250 & 0.251 & <0.001 & -0.161 \\
\rowcolor{gray!20}
manila & 0.728 & 0.731 & <0.001 & -0.511 & 0.890 & 0.762 & <0.001 & 0.995 & 0.877 & 0.776 & <0.001 & 0.977 \\
quito & 0.615 & 0.675 & <0.001 & -0.931 & 0.730 & 0.725 & <0.001 & 0.491 & 0.705 & 0.751 & <0.001 & -0.932 \\
\rowcolor{gray!20}
rome & 0.719 & 0.732 & <0.001 & -0.840 & 0.864 & 0.762 & <0.001 & 0.987 & 0.847 & 0.776 & <0.001 & 0.957 \\
washington & 0.752 & 0.745 & <0.001 & 0.772 & 0.894 & 0.770 & <0.001 & 0.996 & 0.887 & 0.780 & <0.001 & 0.980 \\

            \bottomrule
        \end{tabular}
    }
\end{table}

\fakesection{\rqone}
The first row of Table \ref{tab:table} shows the results for {\bf RQ1}, where 
\queen used ideal simulation when evaluating ensemble fitnesses.
For ensemble sizes $3$, $5$, and $7$, it shows the median fitness of the 
heterogeneous and homogeneous ensembles, and the $p$-value and rank-biserial 
correlation effect size from the two-tailed Mann-Whitney $U$ test.
A $p$-value less than $0.001$ indicates that we can reject the null hypothesis 
of no difference in fitness between heterogeneous and homogeneous ensembles at 
the $99.9\%$ confidence interval.
The rank-biserial correlation effect size ranges from $-1$ to $1$.
Positive values indicate that heterogeneous ensembles tend to be fitter than 
homogeneous ensembles, negative values indicate the opposite trend, and zero 
indicates no difference.
In all three cases, the $p$-values are low enough to reject the null hypothesis
and conclude that there is a difference in fitness.
The effect sizes are all positive and close to $1$, indicating that 
heterogeneous ensembles are almost always fitter than homogeneous ensembles.
Comparing the median fitnesses when the ensemble size is $3$, there is only a 
minor improvement in favor of heterogeneous ensembles ($0.782$ vs. $0.757$).
However, for sizes $5$ and $7$, the improvement appears more substantial 
($0.920$ vs. $0.775$ and $0.920$ vs. $0.783$ respectively).

\fakesection{\rqtwo}
The ten subsequent rows show the results for {\bf RQ2}, where \queen used noisy 
simulation when evaluating ensemble fitnesses.
The $p$-values are low enough to reject the null hypothesis for all three 
ensemble sizes and all ten fake backends except for when the size is $5$ and the 
backend is ``kyoto'' ($p=0.003$).
When the ensemble size is $3$, the effect sizes are now negative in most cases, 
indicating that homogeneous ensembles tend to be fitter than heterogeneous 
ensembles.
However, in the case of $5$ and $7$ circuits, the effect sizes remain positive 
except for when the fake backend is ``kyoto'' or ``quito''.

\section{Related Work}
\label{sec:related}

Schuld \etal \cite{Schuld2018} proposed {\it quantum ensembles} for binary 
classification tasks in the context of quantum machine learning 
\cite{Zhang2020}.
Their aim was to implement classical ensemble learning \cite{Dong2020} in a
quantum setting.
Our aim was to improve the reliability of probabilistic quantum circuits by
evolving them as heterogeneous ensembles and combining their outputs 
classically.
While the problem we used in our evaluation is typically used as a benchmark for 
machine learning classifiers~\cite{Chen2005}, our approach is applicable to any
problem with test cases.
Their concept of a quantum ensemble is implemented within a single circuit, so 
the benefits of redundancy with respect to noise resistance from combining 
diverse circuits (inspired by $n$-version programming \cite{Khoury2012}) does 
not apply, unlike our work.

Rather than using evolutionary algorithms to generate quantum circuits, 
researchers have used them to transform existing circuits.
For example, O'Brien \etal \cite{OBrien2021} introduced a genetic improvement 
approach to transform an arbitrary quantum circuit such that it can be correctly 
and efficiently executed on a specific quantum computer.
Researchers have also applied evolutionary algorithms in a wider context, such
as Muqeet \etal \cite{Muqeet2024b}, who proposed a genetic programming technique 
to generate accurate expression-based quantum noise models.

\section{Conclusion and Future Work}
\label{sec:conclusion}

Our results demonstrate that heterogeneous ensembles of probabilistic quantum 
circuits generally perform better than homogeneous ensembles when using ideal 
simulation.
They also show that heterogeneous ensembles still mostly outperform homogeneous 
ensembles when using noisy simulation, provided the ensembles are large enough.
These findings imply that our approach of evolving heterogeneous ensembles is 
more reliable than evolving individual probabilistic circuits and executing them 
several times.
As future work, we plan to evaluate the impact of explicitly promoting circuit 
diversity within ensembles during evolution.
We also plan on evaluating several different problems and voting mechanisms.

\bibliographystyle{splncs04}
\bibliography{bibliography}

\end{document}